\documentclass[prb,reprint,superscriptaddress,showkeys,floatfix]{revtex4-2}

\usepackage{graphicx}
\usepackage{amsmath,amssymb}
\usepackage{bm}
\usepackage{braket}
\usepackage{upgreek}
\usepackage{hyperref}
\hypersetup
{
	colorlinks=true,
	allcolors=blue,
}
\newcommand{\x}[1] {\mathrm{#1}}
\newcommand{\n} {\x{n}}
\newcommand{\nn} {\x{nn}}
\newcommand{\nnn} {\x{n,nn}}
\newcommand{\lr} {\lambda_\x{R}}
\newcommand{\lpia} {\lambda_\x{PIA}}
\newcommand{\lf} {l_\x{F}}
\newcommand{\tp} {\tau_p}
\newcommand{\tx} {\tau_x}
\newcommand{\tz} {\tau_z}
\newcommand{\lmfp} {l_\x{mfp}}
\newcommand{\lsoc} {l_\x{soc}}
\newcommand{\lrms} {{\lambda}_\x{rms}}
\renewcommand{\Re}{\operatorname{Re}}

\begin{document}

\title{Upper limit of spin relaxation in suspended graphene}

\author{Aron W. Cummings}
\email{aron.cummings@icn2.cat}
\affiliation{Catalan Institute of Nanoscience and Nanotechnology (ICN2), CSIC and BIST, Campus UAB, 08193 Bellaterra, Spain}

\author{Simon M.-M. Dubois}
\affiliation{Institute of Condensed Matter and Nanosciences, Universit\'e catholique de Louvain (UCLouvain), Chemin des \'etoiles 8, B-1348 Louvain-La-Neuve, Belgium}

\author{Pedro Alc\'azar Guerrero}
\affiliation{Catalan Institute of Nanoscience and Nanotechnology (ICN2), CSIC and BIST, Campus UAB, 08193 Bellaterra, Spain}
\affiliation{	Department of Physics, Universitat Autònoma de Barcelona (UAB), Campus UAB, 08193 Bellaterra, Spain}

\author{Jean-Christophe Charlier}
\affiliation{Institute of Condensed Matter and Nanosciences, Universit\'e catholique de Louvain (UCLouvain), Chemin des \'etoiles 8, B-1348 Louvain-La-Neuve, Belgium}

\author{Stephan Roche}
\affiliation{Catalan Institute of Nanoscience and Nanotechnology (ICN2), CSIC and BIST, Campus UAB, 08193 Bellaterra, Spain}
\affiliation{	ICREA -- Instituci\'o Catalana de Recerca i Estudis Avan\c{c}ats, 08010 Barcelona, Spain}

\begin{abstract}
We use a combination of molecular dynamics and quantum transport simulations to investigate the upper limit of spin transport in suspended graphene. We find that thermally-induced atomic-scale corrugations are the dominant factor, limiting spin lifetimes to $\sim$$10$ ns by inducing a strongly-varying local spin-orbit coupling. These extremely short-range corrugations appear even when the height profile appears to be smooth, suggesting they may be present in any graphene device. We discuss our results in the context of experiments, and briefly consider approaches to suppress these short-range corrugations and further enhance spin lifetimes in graphene-based spin devices.
\end{abstract}


\maketitle

\section{Introduction}

Since its first measurement \cite{Novoselov2004}, graphene has proven to be a promising material for a wide range of applications \cite{Ferrari2015, Roche2015, Romagnoli2018, Sierra2021, Yang2022, Roche2024}. In spintronics, graphene is valued for its small spin-orbit coupling (SOC) and hyperfine interaction, both of which suggest that spin lifetimes should be exceptionally long. Indeed, initial theoretical studies predicted lifetimes in the range of microseconds to milliseconds and spin diffusion lengths on the order of tens to hundreds of microns \cite{Ertler2009, Huertas2009}. However, initial measurements found lifetimes orders of magnitude smaller, around 100 ps, with spin diffusion lengths of $1-2$ $\upmu$m \cite{Tombros2007, Han2009, Jozsa2009, Pi2010}. A variety of theories were developed to explain this discrepancy, among them spin-pseudospin coupling \cite{Tuan2014, Cummings2016}, resonant magnetic impurities \cite{Kochan2014}, and substrate effects \cite{Tuan2016}, all of which predict similar lifetimes in dirty or disordered graphene.

After years of improvements in graphene quality, device fabrication, and contact optimization, today's best devices exhibit spin lifetimes reaching $10$ ns \cite{Drogeler2016}, with spins transported over tens of microns \cite{Ingla2015, Drogeler2016, Gebeyehu2019, Panda2020, Bisswanger2022}. These advancements position graphene as an excellent transporter of spin over long distances. However, even in devices where transport is effectively ballistic \cite{Drogeler2016, Bisswanger2022}, these lifetimes remain orders of magnitude smaller than original predictions. Thus, it remains an open question as to what are the factors limiting spin transport in clean graphene.

Here we use a combination of molecular dynamics \cite{lammps_paper, lammps_website} and quantum transport \cite{Fan2021, lsquant_website} simulations to examine spin dynamics and relaxation in suspended graphene. In the absence of extrinsic sources of disorder, corrugations due to thermal fluctuations are the only source of spin relaxation. Our simulations yield spin lifetimes between $1-20$ ns for carrier densities typical of experiments, with the lifetime scaling inversely with Fermi energy. Both the magnitude and scaling are well explained by a theory of spin dynamics in a random Rashba spin-orbit field \cite{Dugaev2011, Zhang2012}.

Prior works examining the effect of corrugation predicted lifetimes in the $\upmu$s to ms range \cite{Fratini2013, Vicent2017}. Those works considered long-wavelength corrugations (tens of nm) that were observed experimentally \cite{Ishigami2007, Locatelli2010}. Meanwhile, our simulations reveal that spin transport is limited by corrugations on the atomic ($<$$1$ nm) scale that may not be revealed by standard probe techniques. This suggests that such ultra-short-range corrugations may be the limiting factor on spin transport in otherwise defect-free graphene. On the experimental side, steps may then be taken to suppress such short-range corrugations and push spin lifetimes beyond the 10-ns scale.

\section{Numerical Methodology}

\subsection{Tight-binding model}

To simulate charge and spin transport in large-area corrugated graphene, we first derive a tight-binding model that captures its electronic and spintronic properties. The bare Hamiltonian for flat suspended graphene is
\begin{gather}
\hat{\mathcal{H}}_0 = \sum\limits_{\left< i,j \right>} c_i^\dagger \, t_{ij}^\n \, c_j + \sum\limits_{\left<\left< i,j \right>\right>} c_i^\dagger \, t_{ij}^\nn \, c_j \label{eq_h0} \\
+ \frac{\x{i}\lambda_\x{I}}{3\sqrt{3}} \sum\limits_{\left<\left< i,j \right>\right>} c_i^\dagger \, \eta_{ij} \hat{s}_z \, c_j, \nonumber \\
t_{ij}^\nnn = t_0^\nnn \cdot \exp\left[ -\beta_\nnn \left( \frac{r_{ij}}{a_\nnn} - 1 \right) \right], \label{eq_hop}
\end{gather}
where $c_i^\dagger$ $(c_i)$ is the creation (annihilation) operator of an electron at carbon site $i$, $t_{ij}^\nnn$ describes hopping between nearest and next-nearest neighbors, $t_0^\nnn$ is the hopping at the equilibrium neighbor distance $a_\nnn$, $r_{ij}$ is the distance between atoms $i$ and $j$, and $\beta_\nnn$ describes the distance dependence of the hoppings. The last term in $\hat{\mathcal{H}}_0$ is the intrinsic SOC of graphene, where $\eta_{ij} = +1(-1)$ for a clockwise (counterclockwise) hopping path between second neighbor sites $i$ and $j$, and $\hat{s}_z$ is the Pauli spin operator for spins perpendicular to the graphene plane.

Equation \eqref{eq_h0} captures local strains in graphene through the distance-dependent hopping, but it does not capture out-of-plane corrugation. For this we introduce a local deformation field,
\begin{equation}
\bm{d}_i = d_i \sigma_i \hat{z}, \label{eq_curv}
\end{equation}
where $d_i$ is the magnitude of the local curvature at site $i$, $\sigma_i$ is its sign ($+/-$ for locally convex/concave curvature), and $\hat{z}$ is the unit vector perpendicular to the graphene plane. We define $d_i$ as the inverse of the radius of the sphere that intersects atom $i$ and its nearest neighbors.

In corrugated graphene, the full Hamiltonian is
\begin{equation}
\hat{\mathcal{H}} = \hat{\mathcal{H}}_0 + \hat{\mathcal{H}}_d,
\end{equation}
where $\hat{\mathcal{H}}_d$ contains the terms induced by the local corrugation,
\begin{align}
\hat{\mathcal{H}}_d &= \mu_d \sum\limits_{i} c_i^\dagger \, d_i \, c_i \label{eq_hd} \\
&+ \beta_\n^d \sum\limits_{\left< i,j \right>} c_i^\dagger \, d_{ij}^\n \, c_j
+ \beta_\nn^d \sum\limits_{\left<\left< i,j \right>\right>} c_i^\dagger \, d_{ij}^\nn \, c_j \nonumber \\
&+ \x{i}\lr^d \sum\limits_{\left< i,j \right>} c_i^\dagger \left( \bm{d}_{ij}^\n \times \bm{n}_{ij} \right) \cdot \hat{\bm{s}} \, c_j \nonumber \\
&+ \x{i}\lpia^d \sum\limits_{\left<\left< i,j \right>\right>} c_i^\dagger \left( \bm{d}_{ij}^\nn \times \bm{n}_{ij} \right) \cdot \hat{\bm{s}} \, c_j \nonumber \\
&+ \x{i}\lambda_\x{I}^d \sum\limits_{\left<\left< i,j \right>\right>} c_i^\dagger \, \eta_{ij}\left( d_{ij}^\nn \right)^2 \hat{s}_z \, c_j . \nonumber
\end{align}
The first term in Eq.\ \eqref{eq_hd} captures charge redistribution induced by the local corrugation. The second line accounts for renormalization of the hopping between neighbors. For nearest-neighbor hopping, $d_{ij}^\n = | \bm{d}_{ij}^\n | \equiv \left| (\bm{d}_i + \bm{d}_j) / 2 \right|$ is the magnitude of the average deformation field of the two neighbors. For second-neighbor hopping, $d_{ij}^\nn  = | \bm{d}_{ij}^\nn | \equiv | \bm{d}_k |$, with $k$ the common neighbor of atoms $i$ and $j$. The third line describes a local Rashba SOC arising from the local curvature, i.e., a local breaking of out-of-plane mirror symmetry. Here, $\bm{n}_{ij}$ is the unit vector pointing from $i$ to $j$ and $\hat{\bm{s}}$ is the vector of Pauli spin operators. The fourth line is a second-neighbor Rashba term, also called a PIA term \cite{Gmitra2013}, with $\bm{n}_{ij}$ the unit vector pointing between second-neighbor sites $i$ and $j$. Finally, the last term is a local intrinsic SOC induced by the local curvature.

\begin{figure}[t]
\centering
\includegraphics[width=\columnwidth]{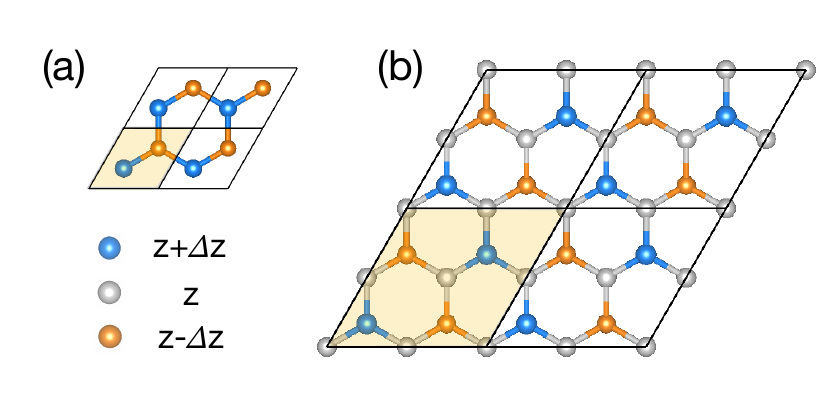}
\caption{Ball-and-stick illustrations of the corrugated graphene samples considered for first-principles simulations, with (a) $1 \times 1$ and (b) $2 \times 2$ unit cells (shaded in orange). Two corrugation amplitudes were included in the fitting process, $\Delta z = 0.075$ and $0.15$ \AA.}
\label{fig_dft_geom}
\end{figure}

All Hamiltonian parameters were obtained by fitting the band structure and spin splitting to first-principles simulations of flat and corrugated graphene. The corrugated graphene samples considered for fitting are illustrated in Fig.\ \ref{fig_dft_geom}.
The first-principles simulations were carried out with the all-electron full-potential linearized augmented plane wave (FP-LAPW) method implemented in the Elk code \cite{elk_code}. The self-consistent calculations with SOC were carried out within the local density approximation with a muffin tin radius of $1.316$ Bohr for carbon atoms and an APW cutoff of $5.32$ Bohr$^{-1}$. A $33 \times 33$ $k$-point mesh was used to sample the first Brillouin zone of pristine graphene, and an equivalent $k$-point density was used for the $2 \times 2$ supercell in Fig.\ \ref{fig_dft_geom}(b).
The fitting of the tight-binding model to the first-principles results was limited to the $[-1, 1]$ eV energy window around the Fermi level; a similar fitting procedure has been used previously in Ref.\ \citenum{Cummings2019} for flat graphene. The values of the fitted parameters are given in Table \ref{tab_params}. As discussed more below, we find that the mean free path and spin lifetime scale as $\sim$$(\lr^d)^{-2}$ and $\sim$$(\mu_d)^{-2}$ respectively. Thus, a factor of 2 variation in these fitting parameters would result in a factor of $\sim$4 variation in the mean free path and spin lifetime.

\begin{table}[t]
\centering
\begin{tabular*}{\columnwidth}{@{\extracolsep{\fill} } c c c}
\hline
Parameter & Description & Value \\
\hline\hline
$t_0^\n$ & 1st-neighbor hopping & $-2.51$ eV \\
$a_\n$ &  & $1.418$ \AA \\
$\beta_\n$ &  & $2.62$ \\
$t_0^\nn$ & 2nd-neighbor hopping & $-0.17$ eV \\
$a_\nn$ &  & $2.456$ \AA \\
$\beta_\nn$ &  & $7.57$ \\
$\lambda_\x{I}$ & intrinsic SOC& $13.4$ $\upmu$eV \\
\hline
$\mu_d$ & local doping & $18.6$ meV$\cdot$nm \\
$\beta_\n^d$ & local hopping & $-4.33$ meV$\cdot$nm \\
$\beta_\nn^d$ & & $-27.8$ meV$\cdot$nm \\
$\lr^d$ & local Rashba& $-9.69$ meV$\cdot$nm \\
$\lpia^d$ & & $-0.322$ meV$\cdot$nm \\
$\lambda_\x{I}^d$ & local intrinsic& $-4.44$ $\upmu$eV$\cdot$nm$^2$ \\
\hline
\end{tabular*}
\caption{Tight-binding parameters for the corrugated graphene Hamiltonian given in Eqs.\ \eqref{eq_h0}, \eqref{eq_hop}, and \eqref{eq_hd}.}
\label{tab_params}
\end{table}

\subsection{Sample generation}

To create corrugated graphene samples we used molecular dynamics simulations, as implemented in the LAMMPS code \cite{lammps_paper, lammps_website}, by applying a finite-temperature thermostat to an initially flat graphene sheet. The flat graphene sheet was first relaxed until reaching a relative energy tolerance of $10^{-8}$ with all forces below $10^{-8}$ eV/\AA. Interatomic forces were computed using the optimized Tersoff and Brenner potentials \cite{Lindsay2010}. We then applied a Nos\'e--Hoover thermostat under the NPT ensemble to thermalize the structure at a desired temperature $T$, with the pressure in all directions fixed to $P=0$. During the initial thermalization step, the thermostat was applied for 100 ps with a time step of 0.1 fs. This is long enough to converge the distributions of nearest-neighbor bond lengths, out-of-plane atomic positions, and local curvatures.

In our transport simulations, we wish to consider a number of different sample configurations at each temperature in order to reduce noise via ensemble averaging. Thus, after the initial thermalization step we applied the thermostat for another 100 ps and saved a snapshot of the atomic positions every 10 ps, generating a total of 11 samples for averaging. This procedure was repeated for $T = 100$, $200$, $300$, and $400$ K.

\subsection{Quantum transport simulations}

To examine charge and spin transport in the corrugated samples, we plug the Hamiltonian of each sample snapshot into a linear-scaling quantum transport (LSQT) approach \cite{Fan2021, lsquant_website}, which enables the efficient simulation of disordered systems containing many millions of atoms. With this approach we calculate the time- and energy-resolved expectation value of the mean-square displacement (MSD) and spin polarization ($\bm{S}$) of an initial quantum state $\ket{\psi}$,
\begin{align}
\x{MSD}(E,t) &= \frac{1}{2} \left[ \Delta X^2 (E,t) + \Delta Y^2 (E,t) \right], \\
\Delta X^2 (E,t) &= \frac{\braket{\psi_X(t) | \delta(E-\hat{\mathcal{H}}) | \psi_X(t)}}{\rho(E)}, \\
\Delta Y^2 (E,t) &= \frac{\braket{\psi_Y(t) | \delta(E-\hat{\mathcal{H}}) | \psi_Y(t)}}{\rho(E)}, \\
\bm{S} (E,t) &= \Re \frac{\braket{\psi(t) | \hat{\bm{s}} \delta(E-\hat{\mathcal{H}}) | \psi(t)}}{\rho(E)},
\end{align}
where $\ket{\psi(t)} = \hat{\mathcal{U}}(t) \ket{\psi}$, $\ket{\psi_X(t)} = [ \hat{\mathcal{X}} , \hat{\mathcal{U}}(t) ] \ket{\psi}$, $\ket{\psi_Y(t)} = [ \hat{\mathcal{Y}} , \hat{\mathcal{U}}(t) ] \ket{\psi}$, $\hat{\mathcal{U}}(t) = \exp(-\x{i}\hat{\mathcal{H}}t/\hbar)$ is the time evolution operator, $\hat{\mathcal{X}}$ ($\hat{\mathcal{Y}}$) is the position operator along the $x$ ($y$) direction, and $\rho(E) = \braket{\psi | \delta(E-\hat{\mathcal{H}}) | \psi}$ is the density of states. To avoid diagonalizing $\hat{\mathcal{H}}$, the operators $\hat{\mathcal{U}}(t)$ and $\delta(E-\hat{\mathcal{H}})$ are expanded as a series of Chebyshev polynomials, while $\Delta X^2$, $\Delta Y^2$, and $\bm{S}$ are reconstructed using the kernel polynomial method \cite{Fan2021, Weisse2006}.

From the mean-square displacement we calculate the time-dependent diffusivity,
\begin{equation}
D(E,t) = \frac{1}{2} \frac{\partial}{\partial t} \x{MSD}(E,t). \label{eq_d}
\end{equation}
From the spin dynamics we calculate the spin lifetime by fitting the time-dependent spin polarization to an exponential decay,
\begin{equation} \label{eq_fit_spin}
S_i (E,t) = \exp \left[ -t / \tau_i(E) \right],
\end{equation}
where $\tau_i$ is the spin lifetime and $i = x,z$ represents the components of spin pointing in or out of the graphene plane, respectively.

For efficient calculation of $\Delta X^2$, $\Delta Y^2$, and $\bm{S}$ over the entire Hamiltonian spectrum, we let the initial state $\ket{\psi}$ be a random-phase state that is initially spin polarized along the $x$-axis or the $z$-axis to study the relaxation of in-plane or out-of-plane spins \cite{Fan2021}. In the transport simulations we consider $560\times560$-nm samples, which are sufficiently large to extract reliable estimates of the mean free path and the spin lifetime, and an expansion of $3000$ Chebyshev polynomials, corresponding to a numerical Gaussian broadening of 26 meV.

\section{Results}

\subsection{Sample structure}

\begin{figure}[t!]
\centering
\includegraphics[width=\columnwidth]{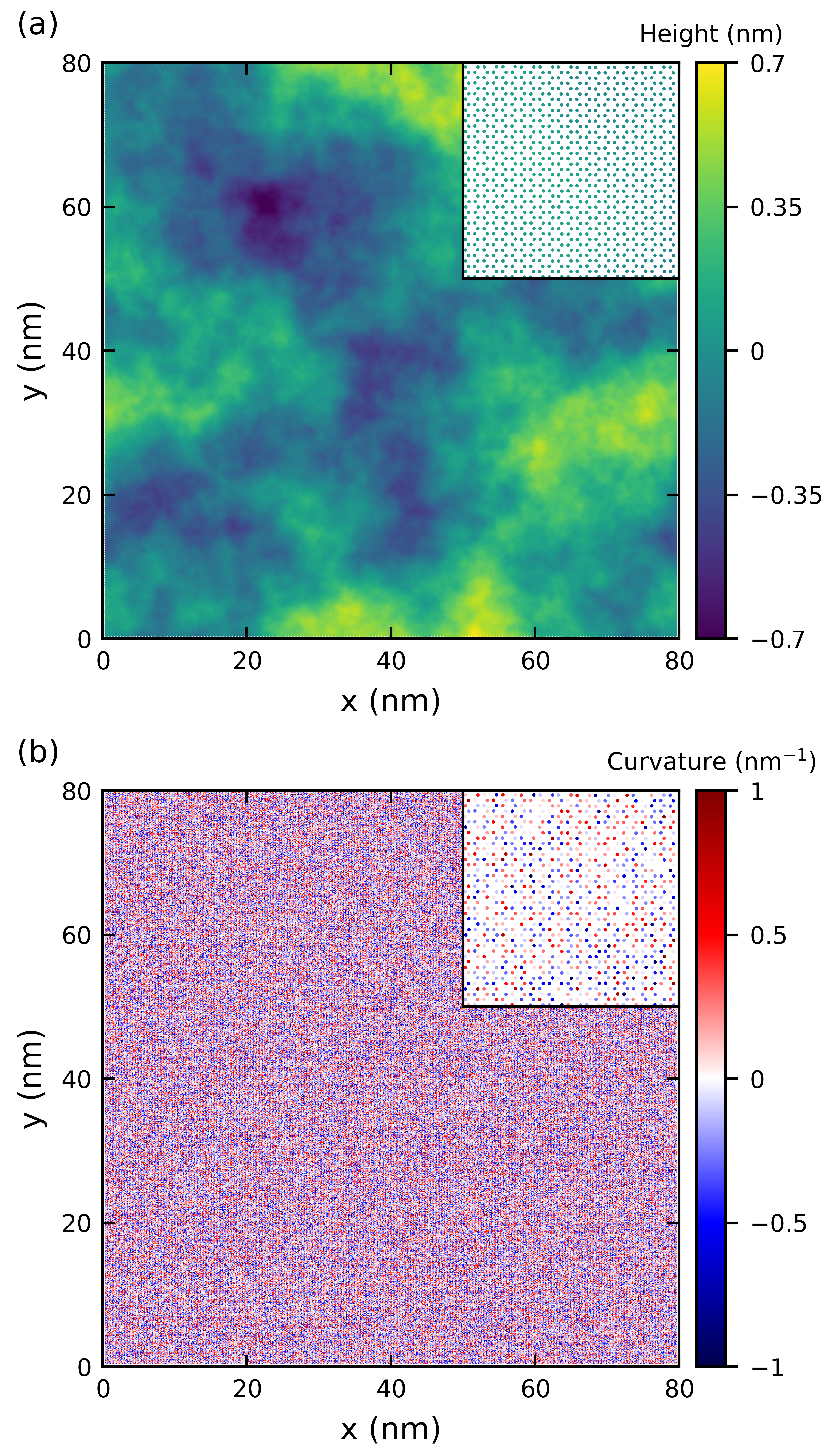}
\caption{Real-space map of the (a) height and (b) local curvature of a graphene sample thermalized at 300 K. The insets show a $5 \times 5$-nm zoom of the lower-left corner of the sample.}
\label{fig_sample}
\end{figure}

First we examine the general structure of the thermally-corrugated graphene samples. Figure \ref{fig_sample}(a) shows the height profile of an $80\times80$-nm sample snapshot thermalized at 300 K. Long-wavelength corrugations on the order of tens of nanometers are evident, with the out-of-plane position varying between $\pm 7$ \AA. This profile is similar to those measured in suspended graphene \cite{Meyer2007, Locatelli2010, Kirilenko2011}.
Figure \ref{fig_sample}(b) shows the local curvature profile of this sample, which is dominated by very short range fluctuations, $<$$1$ nm, with no apparent correlation with the height profile in panel (a). This curvature profile arises from small fluctuations in the height of each atom, superimposed over the longer wavelength height profile. Here we have only shown one sample, but these features are consistent for all samples and temperatures.

From Eq.\ \eqref{eq_hd}, the local SOC is proportional to the local curvature. During transport, electrons will therefore move through a spin-orbit field that looks qualitatively like Fig.\ \ref{fig_sample}(b). As we will see below, this rapidly-varying spin-orbit field will be the limiting factor on spin transport in our simulations.

\subsection{Charge transport}

\begin{figure}[t!]
\centering
\includegraphics[width=\columnwidth]{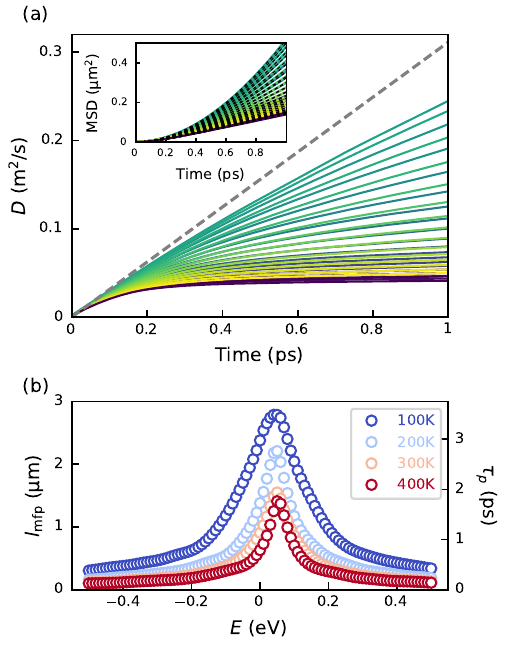}
\caption{Charge transport in thermally-corrugated graphene. (a) Time-dependent diffusivity at 300 K for energies $E \in [-0.5, +0.5]$ eV. The purple (yellow) curves correspond to energies near $-0.5$ ($+0.5$) eV, and the blue/green curves to energies near the Dirac point. The dashed gray line indicates the ballistic limit. Inset: mean square displacement, with the dashed lines showing the fits to Eq.\ \eqref{eq_x2fit}. (b) Mean free path at each temperature. The right y-axis indicates the corresponding momentum relaxation time.}
\label{fig_charge}
\end{figure}

Next we examine the charge transport properties of suspended corrugated graphene. Figure \ref{fig_charge}(a) shows the diffusivity of the 300-K samples for Fermi energies $E \in [-0.5, +0.5]$ eV. The purple (yellow) curves correspond to the highly p-doped (n-doped) regime near $E = -0.5$ $(+0.5)$ eV, and the blue/green curves to energies near the Dirac point. At most energies the diffusivity has not saturated within the simulation time window, especially near the Dirac point, which is indicative of weak scattering and quasi-ballistic transport. For comparison, the dashed gray line indicates the ballistic limit.

To quantify the efficiency of charge transport, we extract the mean free path from our simulations. To do so, the diffusivity must saturate to a constant value $D_\x{sat}$, such that $\lmfp(E) = D_\x{sat}(E) / v(E)$, where $v$ is the Fermi velocity. However, the quasi-ballistic transport indicated in Fig.\ \ref{fig_charge}(a) means that reaching saturation is not achievable within the time and length constraints of our simulations. Thus, to estimate $\lmfp$, we extrapolate our results by fitting the MSD to an expression capturing the transition from the ballistic to the diffusive regime \cite{Li2013, Ornstein1919},
\begin{equation}
\x{MSD} = 2 \lmfp^2 \left( t/\tp - 1 + \x{e}^{-t/\tp} \right),
\label{eq_x2fit}
\end{equation}
where $\tp$ is the momentum relaxation time. The dashed black lines in the inset of Fig.\ \ref{fig_charge}(a) show fits to some of the numerical data.

Figure \ref{fig_charge}(b) shows the fitted $\lmfp$ at each temperature, as well as the corresponding $\tp$ on the right y-axis. The mean free path is long in all cases, from hundreds of nm up to 3 $\upmu$m, approaching measurements in clean hBN-encapsulated graphene \cite{Banszerus2016, Drogeler2016}. Thus, although short-range corrugations appear to be strong, they have only a moderate impact on charge transport. By setting the second line of Eq.\ \eqref{eq_hd} to zero and rerunning our simulations, we find that the mean free path is unaffected, meaning the charge transport properties are determined by the curvature-induced local variations in carrier density, and not by local variations in hopping. We also note that $\lmfp$ decreases with increasing temperature, owing to enhanced corrugation.

\subsection{Spin lifetime}

Next, we examine spin transport in the thermally corrugated graphene samples. From the LSQT simulations we calculate the ensemble spin polarization $\bm{S}(E,t)$, from which we extract the spin lifetime by fitting to Eq.\ \eqref{eq_fit_spin}.

\begin{figure}[t!]
\centering
\includegraphics[width=\columnwidth]{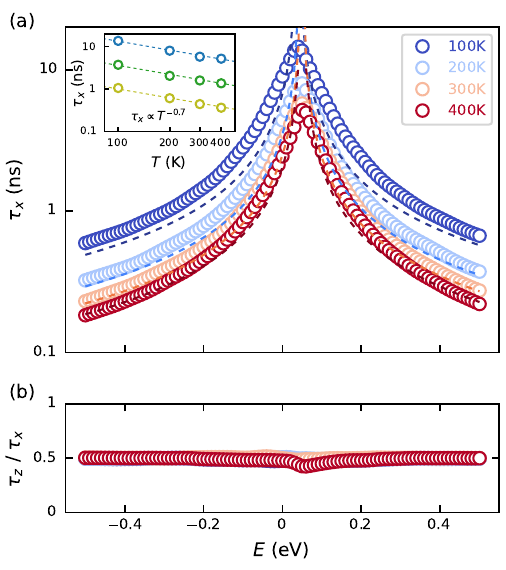}
\caption{Spin relaxation in thermally-corrugated graphene. (a) In-plane spin lifetime, where symbols are numerical simulations and dashed lines are from a theory of spin relaxation in a random Rashba field \cite{Dugaev2011}, see Eq.\ \eqref{eq_tsanal}. Inset: temperature scaling of spin lifetime at the Dirac point, at $E = 0.15$ eV, and at $E = 0.35$ eV (top to bottom). Symbols are numerical simulations and dashed lines show $\tx \propto T^{-0.7}$. (b) Spin lifetime anisotropy vs.\ energy.}
\label{fig_spin}
\end{figure}

Figure \ref{fig_spin}(a) shows the in-plane spin lifetime $\tx$ at each temperature. The symbols are numerical simulations and the dashed lines are from a theory of spin relaxation in a random Rashba field \cite{Dugaev2011}, discussed in more detail below. We first note that $\tx$ monotonically decreases with increasing temperature --- higher temperature yields stronger corrugation and thus stronger spin-orbit coupling. The inset of Fig.\ \ref{fig_spin}(a) shows the scaling of spin lifetime with temperature at the Dirac point, at $E = 0.15$ eV, and at $E = 0.35$ eV (top to bottom). In all cases $\tx \propto T^{-0.7}$, indicated by the dashed lines. The origin of this scaling is discussed in more detail in Section \ref{sec_mechanism} below.

Second, we note the magnitude of the spin lifetime, in the range of $1-20$ ns around the Dirac point. These values are similar to the largest lifetimes obtained experimentally in graphene nonlocal spin valves \cite{Ingla2015, Drogeler2016, Gebeyehu2019, Panda2020}, yet are orders of magnitude lower than theoretical predictions of spin lifetime in graphene with long-range corrugations \cite{Fratini2013, Vicent2017}. This indicates that while short-range corrugations have a relatively weak impact on charge transport, they can serve as a strong limiter of spin transport in otherwise defect-free graphene.

Figure \ref{fig_spin}(b) shows the spin lifetime anisotropy at each temperature. Defined as the ratio $\tz/\tx$, the anisotropy is a useful probe for determining the nature of spin dynamics and relaxation in graphene and graphene heterostructures \cite{Raes2016, Cummings2017}. Here, $\tz/\tx = 1/2$ at all energies and temperatures, indicating that spin relaxation is driven by a D'yakonov--Perel'-like mechanism, with a helical in-plane spin texture arising from the corrugation-induced Rashba SOC \cite{DP1972, Fabian2007}.

We note that to date no experiments have reported $\tz/\tx = 1/2$ in graphene. Measurements of graphene on silicon oxide all indicate isotropic \cite{Raes2016} or nearly isotropic spin relaxation \cite{Raes2017, Ringer2018}. In the latter case the anisotropy was found to be $0.88-0.96$ using oblique spin precession and $0.76-0.82$ using an in-plane magnetic field \cite{Ringer2018}, suggesting spin relaxation driven by a combination of Rashba SOC and extrinsic effects such as magnetic impurities \cite{Kochan2014} and/or contact effects. A theoretical model developed by Zhu and Kawakami demonstrates that finite contact resistance in nonlocal spin valves induces additional spin relaxation that suppresses the measured anisotropy \cite{Zhu2018}. Their closed-form expression provides a way to account for such contact-induced effects, which could be instrumental in revealing the intrinsic anisotropy predicted by our model.

\subsection{Spin relaxation mechanism}
\label{sec_mechanism}

Finally, we examine the spin relaxation mechanism responsible for the features in Fig.\ \ref{fig_spin}. In the semiclassical D'yakonov--Perel' (DP) mechanism, electrons moving through a uniform SOC field undergo spin precession around an effective magnetic field that is perpendicular to their momentum. Randomization of the momentum by charge scattering then leads to random spin precession \cite{DP1972, Fabian2007}. In contrast, here electrons move through a spatially-random SOC field with minimal scattering. From the electron's reference frame these scenarios are equivalent, with both leading to spin relaxation driven by randomized spin precession.

Prior theories have considered spin dynamics and relaxation in a random spin-orbit field \cite{Dugaev2011, Zhang2012, Fratini2013, Vicent2017}. The relevant length scales in this scenario are $\lmfp$; the spin precession length $\lsoc \sim \hbar v / \lrms$, with $\lrms$ the root-mean-square (RMS) spin-orbit strength; the Fermi wavelength $\lf = 2\pi/k$, with $k$ the Fermi wave number; and the length scale of the spin-orbit fluctuations $R$.

When $R$ is the smallest length scale in the system, the in-plane spin lifetime is \cite{Dugaev2011}
\begin{equation}
\tx^{-1} = \pi \left( \frac{\lrms}{\hbar} \right)^2 \left( \frac{R}{v} \right) \pi k R \, .
\label{eq_tsanal}
\end{equation}
This is reminiscent of the standard expression for DP spin relaxation, $\tx^{-1} \sim (\lr/\hbar)^2 \tp$ \cite{DP1972, Fabian2007}. In Eq.\ \eqref{eq_tsanal}, because charge scattering is weak and $R \ll \lmfp$, the time scale of spin-orbit fluctuations is $R/v$ instead of $\tp$.

A distinction between Eq.\ \eqref{eq_tsanal} and the semiclassical DP mechanism is the factor $\pi k R$, a quantum effect that arises when $R \ll \lf$ \cite{Dugaev2011}. In this limit the electron is spread over a wide area and the effective SOC strength is reduced by spatial averaging, yielding slower relaxation for larger $\lf$. As graphene has a linear dispersion, $E = \hbar v k = h v / \lf$, the lifetime is inversely proportional to the Fermi energy.

According to Eq.\ \eqref{eq_tsanal}, the spin lifetime is determined entirely by the geometric properties of the corrugation, via the strength and the length scale of the fluctuating spin-orbit field, $\lrms$ and $R$. To connect Eq.\ \eqref{eq_tsanal} to our simulations, we directly extract these parameters from the graphene samples. Assuming spin relaxation is driven by the corrugation-induced Rashba SOC, the spatially-varying spin-orbit field is $\lambda(\bm{r}) = \lr^d \bm{d}_{ij}^\n \cdot \hat{z}$, where $\bm{r}$ is the set of positions midway between each pair of nearest neighbors. The strength of the SOC field is then $\lrms = \sqrt{\langle \lambda^2(\bm{r}) \rangle}$, with $\langle ... \rangle$ the spatial average over all samples at a given temperature.

We find $R$ by calculating the autocorrelation of the SOC field \cite{Dugaev2011}, $C(\bm{r}-\bm{r}') = \langle \lambda(\bm{r}) \lambda(\bm{r}') \rangle = \mathcal{F}^{-1} \left\{ \mathcal{F}\{ \lambda(\bm{r}) \} \mathcal{F}^*\{ \lambda(\bm{r}) \} \right\}$, where $\mathcal{F}\{...\}$ is the Fourier transform. The autocorrelation is well described by a Lorentzian, $C(\bm{r}-\bm{r}') = \lrms^2 R^2 / \left( |\bm{r}-\bm{r}'|^2 + R^2 \right)$, and we fit to this expression to extract $R$ at each temperature.

Plugging the extracted values of $\lrms$ and $R$ into Eq.\ \eqref{eq_tsanal} yields the dashed lines in the main panel of Fig.\ \ref{fig_spin}(a), which match the numerical simulations very well. This good match indicates that it is the local Rashba SOC that is responsible for spin relaxation, while the local PIA and intrinsic terms play a negligible role. We have verified this in our numerical simulations by setting these terms to zero and finding that the spin lifetime is unchanged.

\begin{figure}[th!]
\centering
\includegraphics[width=\columnwidth]{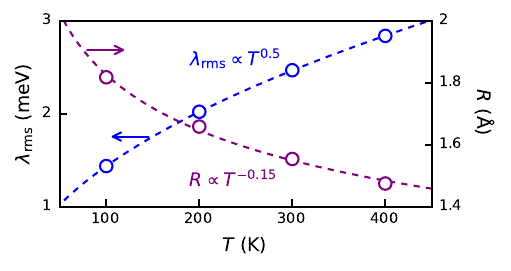}
\caption{Temperature scaling of the RMS Rashba spin-orbit strength (left axis), and the length scale of the spin-orbit fluctuations (right axis). Symbols are numerical results extracted from the corrugated samples, and dashed lines indicate the temperature scaling.}
\label{fig_scaling}
\end{figure}

Figure \ref{fig_scaling} shows the temperature dependence of $\lrms$ and $R$. We see that $R \approx 1.5-1.8$ \AA~and scales weakly with temperature as $T^{-0.15}$. These values are only slightly larger than the distance between nearest neighbors, highlighting the atomic-scale range of the corrugations. Meanwhile, $\lrms \propto T^{0.5}$, yielding $\tx \propto 1 / (\lrms R)^{2} \propto T^{-0.7}$. Finally, we note that $\lrms$ is quite large for graphene, between $1-3$ meV. These large values of local SOC, arising from the extremely short range of the corrugations, are what ultimately limit the spin lifetime to the nanosecond range.

\section{Summary and Discussion}

We have used a combination of molecular dynamics and quantum transport simulations to investigate the upper limit of spin transport in suspended graphene. We find that atomic-scale corrugations are the dominant factor, limiting spin lifetimes to $\sim$$10$ ns by inducing a strongly-varying local spin-orbit coupling.

We stress that the only difference with prior results predicting micro- to millisecond lifetimes is the magnitude and length scale of the corrugations considered. Prior studies \cite{Huertas2009, Fratini2013, Vicent2017} were based on measurements of the height profile \cite{Ishigami2007, Locatelli2010}, which fluctuated over tens of nanometers, similar to Fig.\ \ref{fig_sample}(a). If we had used a similar curvature profile, our simulations would also yield microsecond lifetimes.

However, we employed an atomically local definition of curvature, Eq.\ \eqref{eq_curv}. This reveals strong local variations of corrugation, even if the height profile looks relatively smooth. This definition requires precise knowledge of the positions of each atom and its neighbors, which is available in molecular dynamics simulations but would be extremely challenging to access experimentally, making it perhaps a ``hidden'' feature limiting spin transport in graphene.

The question therefore arises: is there evidence of corrugation-limited spin lifetime in experiments? If so, one would expect improved lifetimes in systems with reduced corrugation. A substrate like SiO$_2$ may suppress corrugations compared to suspended graphene, as measurements generally reveal surface roughness similar to or slightly lower than what we find in our MD simulations \cite{Lui2009, Dean2010, Toksumakov2023}. We note that this is surface roughness and not local curvature, but using this as a proxy, we might expect upper bounds in graphene/SiO$_2$ that are perhaps a few times larger than our simulations. Today the best measurements of graphene on SiO$_2$ yield lifetimes of $2-4$ ns \cite{Gebeyehu2019, Panda2020}.

Encapsulation in hBN should further reduce roughness, with the added benefit of protecting graphene from contaminants. High-mobility measurements reveal lifetimes up to $3$ ns \cite{Ingla2015}. However, ferromagnetic contacts may be suppressing the lifetime in such devices, and it is unclear if the upper limit has been reached. This is also true of the best measurement to date \cite{Drogeler2016}, where contacts were pre-fabricated on SiO$_2$ and graphene was placed on top by an hBN stamp, avoiding any exposure to processing. The resulting lifetime of $12$ ns is promising but may also be contact-limited, as additional effects can appear in nonlocal spin valves when transport is ballistic \cite{Vila2020, Drogeler2017}. Measurements of graphene on hBN yield surface roughness that is up to $6\times$ smaller than our simulations \cite{Dean2010, Toksumakov2023}, suggesting an upper limit of spin relaxation one to two orders of magnitude larger than what we find for suspended graphene.

A signature of our results is the maximal spin lifetime at the Dirac point, which then decays as $1/E$. In contrast, the above-mentioned measurements exhibit minimal lifetime at the Dirac point, or at least relatively flat energy dependence. This suggests that corrugation may not yet be a limiting factor. Meanwhile, recent work claiming corrugation-limited spin transport found a maximal nonlocal spin signal around the Dirac point \cite{Zhou2024}, but this was not converted to a density-dependent lifetime. This work also reported unusual temperature scaling, with a maximum lifetime of $\sim$$5$ ns at 150 K that decreased for lower and higher temperatures \cite{Zhou2024}. This contrasts with our numerical results, but may arise from an interplay between corrugations and the semi-suspended/pinned nature of their samples.

Overall, clear trends connecting corrugation to measured spin lifetimes in single-layer graphene are difficult to see, as extrinsic effects may still be dominant. However, insight may be gained by comparing to multilayered graphene structures that should naturally exhibit weaker corrugation. For example, lifetimes in hBN-encapsulated bilayer graphene have reached beyond $5$ ns \cite{Bisswanger2022} and up to $8$ ns \cite{Xu2018}, positioning it above most cases of single-layer graphene. Measurements of epitaxial graphene on SiC estimated a spin diffusion length of $\sim$$150$ $\upmu$m and a lifetime above $100$ ns \cite{Dlubak2012}, one order of magnitude higher than the best measurements of single-layer graphene \cite{Ingla2015, Drogeler2016, Gebeyehu2019, Panda2020}. Finally, recent measurements of graphite also found spin lifetimes in excess of $100$ ns \cite{Markus2023}.

The quality of the measurements in layered graphene materials seems to indicate the benefit of reducing corrugation, but may also arise from the protection from contaminants afforded by layering, or from improved impurity screening. Looking ahead, other approaches to reducing corrugation, such as strain, may be promising to move spin lifetimes in single-layer graphene above the 10-ns regime. Our findings also predict some features specific to suspended graphene which could be probed experimentally, such as the temperature and energy dependence of the spin lifetime, as well as the anisotropy, which may be probed with combined experimental \cite{Ringer2018} and theoretical \cite{Zhu2018} analysis.

\begin{acknowledgments}
We thank E.Y.\ Sherman for insightful discussions.
ICN2 is funded by the CERCA Programme / Generalitat de Catalunya and supported by the Severo Ochoa Centres of Excellence programme, Grant CEX2021-001214-S, funded by MCIN / AEI / 10.13039.501100011033.
A.W.C., P.A.G., and S.R.\ acknowledge support from the Ministerio de Ciencia e Innovación (MCIN) under grant no.\ PID2019-106684GB-I00 financed by MCIN / AEI / 10.13039/501100011033, and support from MCIN with European funds‐NextGenerationEU (PRTR‐C17.I1) and 2021 SGR 00997, funded by Generalitat de Catalunya.
Simulations carried out by ICN2 were performed at the Center for Nanoscale Materials, a U.S.\ Department of Energy Office of Science User Facility, supported by the U.S.\ DOE, Office of Basic Energy Sciences, under Contract No.\ 83336.
S.M.-M.D.\ and J.-C.C.\ acknowledge financial support from the European Union's Horizon 2020 Research Project and Innovation Program - Graphene Flagship Core3 (No.\ 881603), from the F\'ed\'eration Wallonie-Bruxelles through the ARC Grant ``DREAMS'' (No.\ 21/26-116), from the EOS project ``CONNECT'' (No.\ 40007563), from the EU Pathfinder project ``FLATS'' (No.\ 101099139), and from the Belgium F.R.S.-FNRS through research project No.\ T.029.22F.
Computational resources have been provided by the supercomputing facilities of UCLouvain (CISM) and the Consortium des \'Equipements de Calcul Intensif (C\'ECI), funded by the F.R.S.-FNRS (No.\ 2.5020.11) and by the Walloon Region.
\end{acknowledgments}

\bibliography{manuscript_graphene_spin_suspended_arxiv}

\end{document}